\magnification\magstep1
\baselineskip=14pt
\input psbox
\let\fillinggrid=\relax
\def\no{\noindent}
\def\a{\alpha}
\def\b{\beta}
\def\g{\gamma}

\def\se{s_z}
\def\s{\sigma}
\def\dm{{h_1}}

\def\bm{M'}
\def\be{\eta+\omega}
\def\om{\omega}
\def\bPhi{{\bar \Phi}}
\def\bPsi{{\bar \Psi}}
\def\bchi{{\bar \chi}}

\no
\centerline{\bf Quantum Hall Transition in an Array of Quantum Dots}
\vskip 0.2in
\centerline{K. Ziegler}
\vskip 0.3in
\centerline{Max-Planck-Institut f\"ur Physik Komplexer Systeme}
\centerline{Au\ss enstelle Stuttgart, Postfach 800665, D-70506 Stuttgart,
Germany}
\bigskip
\vskip 0.6in
\no 
Abstract:\par
\no
A two-dimensional array of quantum dots in a magnetic field is considered.
The electrons in the quantum dots are described as unitary
random matrix ensembles. The strength of the magnetic field is such that
there is half a flux quantum per plaquette. This model exhibits
the Integer Quantum Hall Effect. For $N$ electronic states per
quantum dot the limit $N\to\infty$ can be solved by a saddle point
integration of a supersymmetric field theory.
The effect of level statistics on the density of states and the Hall
conductivity is compared with the effect of temperature fluctuations.
\bigskip
\bigskip
\no
We consider a two-dimensional array of quantum dots in a homogeneous
magnetic field perpendicular to the array.
A quantum dot in an array is a complex finite system of electrons, subject
to strong Coulomb interaction and a confining potential. Even if the number of
electrons is small there is a large number of electronic states in a given
energy interval. Therefore,
we are forced to use a statistical description of the quantum dot. A typical
feature of such a complicated non-integrable system is level repulsion.
The latter, also found in other complex many-particle systems like atomic
nuclei [1],
atoms [2] or metallic particles [3], can be conveniently described by random
matrix ensembles [4]. Since the magnetic field breaks the time--reversal
invariance in the dot, an appropriate model is the Gaussian unitary ensemble
(GUE).
Electrons can travel in the array of quantum dots due to tunneling between
neighboring dots. On the square--array, which will be considered in this
article, the
tunneling rates are $t$ and $t'$ for nearest and next nearest neighbors,
respectively (cf. Fig.1). The coupling between the individual quantum
dots due to these tunneling processes is weak. This allows us to assume
that the statistical occupation of the electronic states in each dot is
uncorrelated between different dots. Thus the quantum dots can be
represented by independent random matrix ensembles. Moreover, we also assume
for simplicity that the tunneling processes are independent, i.e., the
tunneling electrons do not interact with each other. 

For very weak tunneling rates the array should behave like an insulator
because of the fluctuations of the energy levels. One
would expect for increasing tunneling rates that a metallic regime can be
reached where the array becomes conducting. However, due to the 
statistical fluctuations of the energy levels
in the dots the effect of Anderson localization must play a crucial role
in the array. Anderson localization prevents
a two-dimensional system to become metallic, at least if no or only a weak
magnetic field is present [5]. On the other hand, in
the two-dimensional electron gas in a homogeneous magnetic field
quantum Hall transitions (QHT) have been observed which are
accompanied by delocalized electronic states [6]. A QHT occurs if a gap opens
in a band of electronic states. This phenomenon is known, for instance,
from electrons which
are subject to a homogeneous magnetic field {\it and} a periodic potential
[7].
Depending on the magnetic field the electrons form several subbands where
each subband contributes $e^2/h$ to the Hall conductivity [8,9,10]. As an
approximation of the periodic potential one can use a tight binding model
where the lattice constant is given by the period of the potential.
In this article we will study the effect of the statistics of energy levels
and the effect of thermal fluctuations on the QHT.

There are two different approaches to the transport in quantum dots.
One is based on the S-matrix, the other one on the Hamiltonian. The former
is very useful for numerical simulations because it describes directly the
reflection and transmission of the electrons through the quantum dots [11,12].
The latter, however, requires the application of
linear response theory to get a conductivity via Kubo's formula.
In this article the Hamiltonian representation will be used.
The effective Hamiltonian of an array of quantum dots reads as a quadratic
form $\sum{\hat H}_{r,r'}^{\a,\a'}c_{r}^\a c^{\a'\dagger}_{r'}$
in the fermion creation and annihilation operators $c^\dagger$, $c$
with the matrix elements
$$
{\hat H}_{r,r'}^{\a,\a'}=H_{r}^{\a,\a'}\delta_{r,r'}+H'_{r,r'}
\delta_{\a,\a'}+V_r\delta_{r,r'}\delta_{\a,\a'},
\eqno (1)$$
where $\a,\a'=1,...,N$ label the $N$ electronic states in the quantum dots
and $r$ and $r'$ label positions of the quantum dots in the two-dimensional
array.
In general, tunneling between all $N$ states should be allowed with some
probability, depending exponentially on the energy of the states $\a$ and
$\a'$. To include this would require a
detailed knowledge of these states. Therefore, we choose as a simplifying
approximation the assumption that there is tunneling only between states
with the same
$\a$ at nearest or next nearest neighbor dots with fixed tunneling rates.
The distance between neighboring dots is measured in units of
$(\phi_0/2B)^{1/2}$. Typical distances are $a=100...500 nm$ [13]. The
magnetic field for the creation of one flux quantum per
plaquette is $B=\phi_0/a^2\approx 0.016...0.4T$. This regime is
accessible in natural crystals ($a\approx 0.5 nm$) only with astronomical
magnetic fields.

The electron can occupy statistically states inside the
quantum dot which are represented by the matrix elements $H^{\a,\a'}_r$.
$H$ is the $N\times N$ Hermitean Hamiltonian ($H^\dagger=H$) of
a quantum dot with $N^2$ statistically independent matrix
elements. They are Gaussian distributed with zero mean and
$\langle H_r^{\alpha \alpha'} H_{r'}^{\alpha''\alpha'''}
\rangle =(g/N)\delta^{\alpha\alpha'''}\delta^{\alpha'\alpha''}\delta_{r,r'}$.
$g$ is the strength of the level fluctuations. This depends on the strength
of the interaction beween the electrons inside the dot. Therefore, $g$
increases with the number of electrons per dot and with increasing
confinement.

The tunneling is represented by the Hamiltonian $H'$.
This reads in Landau gauge (with $r=(x,y)$) for flux $\phi$ per plaquette
$$
H'_{r,r'}=te^{2i\pi y\phi/\phi_0}\delta_{r',r+e_x}+t\delta_{r',r+e_y}
\pm it'e^{2i\pi y\phi/\phi_0}\delta_{r',r+e_x\pm e_y}+h.c.
\eqno (2)$$
For the special case of half a flux quantum per plaquette ($\phi=\phi_0/2$)
the phase factor in (2) is real and changes only sign between nearest
neighbors in $y$--direction.
Finally, the potential term $V_r$ represents an additional external
(e.g. electric) field. Here we regard a staggered chemical potential
$V_r=(-1)^{x+y}\mu$ which opens a gap $2\mu$ in the spectrum of the electrons,
as will be explained below [14]. It is probably difficult to implement a
staggered potential in a real sample of quantum dots. However, the parameter
$\mu$ plays only the role of a gap which could be created also by other
means.

We choose for the tunneling rate $t=1$.
Therefore, $\mu$, $t'$ are measured in units of $t$ and $g$ is measured in
units of $t^2$.
If we identify fermions with the four corners of the unit cell (Fig.1)
the tunneling matrix $H'$ can be diagonalized by a Fourier transformation.
This gives a $4\times4$ matrix in Fourier space.
$H$, the Hamiltonian of a dot, is a diagonal matrix with respect
to the four corners in the sublattice representation 
$H=(H_{1}^{\a,\a'},H_{2}^{\a,\a'},H_3^{\a,\a'},H_{4}^{\a,\a'})$.
A similar model with correlated randomness $H_1^{\a,\a'}=H_3^{\a,\a'}
=-H_2^{\a,\a'}=-H_4^{\a,\a'}$ was considered in Ref. [15].

We begin the discussion of the model with the analysis of an array where
the interaction of the electrons inside the quantum dots are
neglected. It can be understood as a tight-binding model for 
non-interacting electrons in
a metall with some electronic bands in a magnetic field [14,16].
The Fourier components of $H'$ can be expanded around the four nodes
$k=(\pm\pi,\pm\pi)$ for $k=(\pm\pi,\pm\pi)+ap$ with small $p$ vectors.
After a global orthogonal transformation
the Hamiltonian reads
$$
H''(p)=2\pmatrix{
\mu-t'&ip_x-p_y&-2t'(p_x+p_y)&0\cr
-ip_x-p_y&-\mu+t'&0&-2t'(p_x-p_y)\cr
-2t'(p_x+p_y)&0&\mu+t'&p_y+ip_x\cr
0&-2t'(p_x-p_y)&p_y-ip_x&-\mu-t'\cr
}
$$
$$
\equiv\pmatrix{
H''_{11}&H''_{12}\cr
H''_{21}&H''_{22}\cr
}.
\eqno (3)$$
The last equation combines the $4\times 4$--structure to a
$2\times 2$--structure with $2\times2$ block matrices $H''_{ij}$.
Neglecting terms $O(p^2)$ the Green's function $({\hat H}+i\omega)^{-1}$
decays into a diagonal block structure
$${\hat G}(i\omega)=\pmatrix{
(H''_{11}{\bf 1}_N+h_1+i\omega
)^{-1}&0\cr
0&(H''_{22}{\bf 1}_N+h_1+i\omega
)^{-1}\cr
}
\eqno (4)$$
with the diagonal matrix $h_1=(H_1+H_3,H_2+H_4)$.
Thus the diagonal elements are statistically independent.
${\bf 1}_N$ is the $N\times N$--unit matrix.
It is interesting to notice that the matrices $H''_{jj}=m_j\sigma_z+
i\nabla_x\sigma_x\mp i\nabla_y\sigma_y$ represent two independent
two-dimensional Dirac
Hamiltonians with masses $m_1=\mu-t'$ and $m_2=\mu+t'$, respectively.

The current density in a Dirac model can be calculated from the response
to an external vector potential $q_y$ [17]. The introduction of
this vector potential is equivalent to a change of the boundary conditions,
a concept extensively used in numerical investigations of Anderson
localization [18]. The response to the vector potential leads
to the Hall conductivity $\sigma_{xy}$ in terms of Green's
functions. We obtain for $q_y\sim0$ the expression [14,15]
$$
\sigma_{xy}\approx i{\sum_{r,r',r''}}'\int Tr[\sigma_x{\hat G}_{rr'}(E-i\om)
{\hat G}_{r'r''}(E-i\om)\sigma_y{\hat G}_{r''r}(E-i\om)]{d\omega\over2\pi}.
\eqno (5)$$
Here $\sum'$ is the sum normalized with the number of quantum dots in the array
and the number of energy levels $N$.
If there is only one electron per dot the energy spectrum has discrete
levels which are well--separated.  For instance, with a harmonic oscillator
potential
for the dot we have $E_n=\hbar\omega_p(n+1/2)$. The separation of the energy
levels in the single electron case allows us to neglect all levels with $n>0$.
Consequently, there is no statistics of energy levels and we can write $\dm=0$.
For the Hall conductivity in units of $e^2/h$ we find
$$
\sigma_{xy}=(1/2)[{\rm sign}(m_1)\Theta(|m_1|-|E|)
+{\rm sign}(m_2)\Theta(|m_2|-|E|)],
\eqno (6)$$
where $\Theta$ is the Heaviside step function.
This result reflects correctly the qualitative behavior of the Hall
conductivity at the QHT:
The Hall conductivity of the original
lattice fermion problem is the sum of the Hall conductivities from
the light Dirac mass ($m_1$) and the heavy Dirac mass ($m_2$), such that the
total
$\sigma_{xy}$ has a jump from 0 to 1 if the light mass changes the sign
(i.e., exchange of particles and holes in the Dirac model).
Thus the Dirac fermions, together with the Hall conductivity of Eq. (5),
represent a simple picture for a Hall transition.
Special cases are $\mu=0$ which gives $\sigma_{xy}=0$ and the (unrealistic)
case $t'=0$ with $\sigma_{xy}=({\rm sign}(\mu)/2\pi)\Theta(|m_1|-|E|)$.
The sharp step--like QHT is only possible in an ideal systems of
non-interacting lattice electrons at zero temperature. In order to compare
with real systems we have to include the statistical fluctuations of the
energy levels as well as thermal fluctuations. The latter are taken into
account by replacing the integral over $\omega$ in (5) by a summation
over discrete Matsubara frequencies $\omega_n=(2n+1)\pi T$
($n=0,\pm1,\pm2,...$).
This leads to a thermal broadening of the step--like behavior of $\sigma_{xy}$.

The effect of the level fluctuations is evaluated by
averaging $\sigma_{xy}$ over the random matrix elements of $\dm$. In order to
derive a simple expression for the limit of infinitely many energy levels
per dot ($N\to\infty$) it is convenient to write
the product of Green's functions $G=(H_0+h_1+z\sigma_0)^{-1}$
($H_0$ is either $H''_{11}$ or $H''_{22}$)
in the expression of the Hall conductivity formally as a functional integral
of a supersymmetric model [15,19,20]
$$
G_{rr'}^{\a\a'}G_{r'r''}^{\b\b'}G_{r''r}^{\g\a}=
\langle\bPsi_r^\a\Psi_{r'}^{\a'}\bchi_{r'}^\b\chi_{r''}^{\b'}
\bPsi_{r''}^\g\Psi_r^\a\rangle_S
-\langle\bchi_r^\a\chi_{r'}^{\a'}\bPsi_{r'}^\b\Psi_{r''}^{\b'}
\bchi_{r''}^\g\chi_r^\a\rangle_S
\eqno (7)$$
with $\langle ...\rangle_S=\int ... \exp(-S_1)\prod_r d\Phi_r d{\bar \Phi_r}$
and with the supersymmetric action (sum convention for $\a$)
$$
S_1=-i\se\sum_{r,r'\mu,j,j'}\Phi_{r,\mu,j}^\a(H_0+z\s_0)_{r,j;r',j'}
\bPhi_{r',\mu,j'}^\a
-i\se\sum_{r,\mu,j}(\Phi_{r,\mu,j}^{\a'}\dm_r^{\a'\a}\bPhi_{r,\mu,j}^{\a}),
\eqno (8)$$
where $\se ={\rm sign}({\rm Im}z)$ and the field
$\Phi_{r,j}^\a=(\Psi_{r,j}^\a,\chi_{r,j}^\a)$.
The first component is Grassmann and the second complex.
$\mu=1,2$ labels the complex and the Grassmann components, and $j=1,2$
labels the two components of the Dirac model. This choice guarantees a
normalized functional.
Consequently, the averaging with respect to the Gaussian distributed 
matrix elements of $\dm$ can be performed in the functional integral
as $\langle\exp(-S_1)\rangle_{h_1}=\exp(-S_2)$ with
the effective action $S_2$. The latter is obtained from $S_1$ by replacing
the second term with $(g/N)\sum_{r,\mu,j}(\Phi_{r,\mu,j}^\a
\bPhi_{r,\mu,j}^\a)^2$.
Thus we have derived an effective field theory for $\Phi$ which serves as
a generating functional for the average product of Green's functions.
It is important to notice that {\it not only} $\dm$ creates the
interaction in $S_2$ but also other types of random terms in $S_1$.
For instance, the interaction can also be created by a term
$(N/g)(Q_{r,\mu,j})^2 -2i\se Q_{r,\mu,j}\Phi_{r,\mu,j}^{\a}
\bPhi_{r,\mu,j}^{\a}$ as the second term in $S_1$, followed by an integration
over the matrix field $Q$. This field, in contrast to the random matrix $\dm$,
does not depend on the index $\a$ of the electronic states inside the quantum
dot. This means that the distribution $\dm$ can be transformed into another
distribution with a new `random variable' $Q$ (which does not have a
probability measure but some generalized distribution including Grassmann
variables). In other words, we can write, after integrating out the field
$\Phi$,
$$
\langle[(H_0+\dm +z\s_0)^{-1}]^{\a\a}...\rangle_{\dm}=
\langle[(H_0+2Q+z\s_0)^{-1}]^{\a\a}...\rangle_Q.
\eqno (9)$$ 
The distribution which belongs to $\langle ...\rangle_Q$ was investigated
in detail in [20]. Here we need only the result for leading order in
$N$:
$\langle ... \rangle_Q =\int ...\exp(-NS(Q,P))\prod_r dP_r dQ_r$
with diagonal matrix fields $Q_r$, $P_r$ and
$$
S(Q,P)={1\over g}\sum_r[Tr(Q_{r}^2)+Tr(P_{r}^2)]
$$
$$
+\log\det(H_0+2Q+z\s_0)-\log\det(H_0-2iP+z\s_0)
\eqno (10)$$
The number of levels $N$ appears in front of the action.
Thus the effect of the statistics of the energy levels can be
evaluated for $N\to\infty$ in saddle point (SP) approximation. The SP equation
reads
$$
{\delta\over\delta Q}\big\lbrack {1\over g}Tr(Q_{r}^2)+
\log\det(H_0+2Q+z\s_0)\big\rbrack=0.
\eqno (11)$$
A second SP equation appears from the variation of $P$ by
replacing $Q\to -iP$. As an ansatz we take a uniform SP solution
$Q_0 =-i P_0 =(1/2)[i\eta\s_0+M_s\s_3]$.
Then (11) leads to the conditions $\eta =(\be-iE) g I$, $M_s=-m_1gI/(1 +gI)$
with the integral $I=\int\lbrack (m_1+M_s)^2 +(\be-iE)^2+k^2\rbrack^{-1}d^2k/2
\pi^2$. This result means that disorder shifts the frequency $\om\to\om+\eta$
and the Dirac mass $m_1\to\bm=m_1+M_s$,
where $\eta(m_1,\om)$ and $M_s(m_1,\om)$ are solutions of the SP equation.
For instance, with $\omega=0$ we have
$\eta^2=(1/4)(M_c^2-m_1^2)\Theta(M_c^2-m_1^2)$ where $M_c=2e^{-\pi/g}$.
The sign of $\eta$ is fixed by the condition that $\eta$ must be analytic in
$\om$. This implies ${\rm sign}(\eta)={\rm sign}(\om)$.
The average density of states (DOS)
is proportional to $\eta$ in the $N\to\infty$--limit.
Thus we have a narrow DOS for the array of quantum dots of width $2M_c$
in contrast to the isolated dot which has a semicircular density
of width $2\sqrt{g}$.
The DOS vanishes for $E=0$ in the absence of level fluctuations. The
creation of a non--zero DOS due to level fluctuations is a non--perturbative
effect.

At $T=0$ and $E=0$ the Hall conductivity per fermion level reads in the
limit $N\to\infty$ and with the approximation that $\bm$ and $\eta$ do not
depend on $\om$
$$
\s_{xy}\approx1/2+{\rm sign}(m_1)\Big[1/2-(1/\pi)
{\rm arctan}(\sqrt{M_c^2/m_1^2-1})\Theta(M_c^2-m_1^2)\Big].
\eqno (12)$$
The Hall conductivities are plotted in Fig.2 for $T=0.1$ with and
without level fluctuations.
It is remarkable that the Hall conductivity is enhanced by the
level fluctuations for $\sigma_{xy}<1/2$ whereas it is suppressed for
$\sigma_{xy}>1/2$. The effect of these fluctuations is strictly constrained to
the interval $2M_c$.

{\it Conclusions}
In a square--array of quantum dots with $N$ electronic states per dot
we have investigated the DOS and the Hall conductivity. Both quantities
are significantly affected by the statistical fluctuations of the energy
levels. In particular, the Hall conductivity, which is step--like
at the QHT in the absence of fluctuations, has a more complicated behavior
in the presence of
level fluctuations. Thermal fluctuations have a different effect on the
Hall conductivity; they lead to a simple broadening of the step--like
behavior.

Only the average quantities have been considered.
However, it is possible within the same method described in this article
to study also higher moments of these quantities.
\vfill
\eject
\no
{\bf References}
\bigskip
\no
[1] E.P. Wigner, Ann.Math.{\bf 53}, 36 (1951), ibid. {\bf 67}, 325 (1958)

\no
[2] N. Rosenzweig and C.E. Porter, Phys.Rev. {\bf 120}, 1698 (1960)

\no
[3] L.P. Gorkov and G.M. Eliashberg, JETP {\bf 21}, 940 (1965)

\no
[4] M.L. Mehta, {\sl Random matrices} (Academic Press, New York, 1967) 

\no
[5] P.A. Lee and T.V. Ramakrishnan, Rev. Mod. Phys. {\bf 57}, 287 (1985)

\no
[6] K. von Klitzing, G. Dorda and M. Pepper, Phys.Rev.Lett. {\bf 45}, 494
 (1980)

\no
[7] R.D. Hofstadter, Phys.Rev. {\bf B 14}, 2239 (1976)

\no
[8] D. Thouless, {\sl The Quantum Hall Effect}, edited by R.E. Prange and
S.M. Girvin

{\ }(Springer-Verlag, New York, 1990)

\no
[9] D. Pfannkuche and R.R. Gerhardts, Phys.Rev. {\bf B 46}, 12 606 (1992)

\no
[10] B.L. Johnson and G. Kirczenow, Phys.Rev.Lett. {\bf 69}, 672 (1992)

\no
[11] U.H. Baranger and P.A. Mello, Phys.Rev.Lett. {\bf 73}, 142 (1994) 

\no
[12] R.A. Jalabert, J.-L. Pichard and C.W.J. Beenakker, Europhys.Lett.
{\bf 27}, 255 (1994)

\no
[13] D. Weiss, {\sl Elektronen in ``k\"unstlichen'' Kristallen} (Verlag
Harri Deutsch, 

{\ }Frankfurt, 1994)

\no
[14] A.W.W. Ludwig, M.P.A. Fisher, R. Shankar and G. Grinstein, Phys.Rev.{\bf
B50}, 7526 

{\ }(1994)

\no
[15] K. Ziegler, Europhys.Lett.{\bf 28 } (1), 49 (1994)

\no
[16] M.P.A. Fisher and E. Fradkin, Nucl.Phys.B251[FS13], 457 (1985)

\no
[17] J.D. Bjorken and S.D. Drell, {\sl Relativistic Quantum Mechanics}
(Mc Graw Hill,

{\ }New York, 1964)


\no
[18] D.J. Thouless, Phys.Rep. {\bf 13}, 93 (1974)

\no
[19] K.B. Efetov, Adv. in Phys. {\bf 32}, 53 (1983)

\no
[20] K. Ziegler, Nucl.Phys. {\bf B344}, 499 (1990)

\vfill
\eject
\no
{\bf Figure Captions}
\bigskip
\bigskip
\no
Fig.1: Schematic picture of an array of quantum dots with nearest ($t$) and
next nearest neighbor ($t'$) tunneling. The square denotes the unit cell
of the translational invariant array with magnetic flux $\Phi=\Phi_0/2$.
\bigskip
\no
Fig.2: Hall conductivity $\sigma_{xy}$ in units of $e^2/h$  as a function
of the effective chemical potential $m=\mu-t'$ at temperature $T=0.1$.
The circles are without level fluctuations and the full curve is with
level fluctuations with variance $g=1.36$.
\vfill
\eject
\centerline{Fig.1:} 
$$\psannotate{\psboxto(5cm;5cm){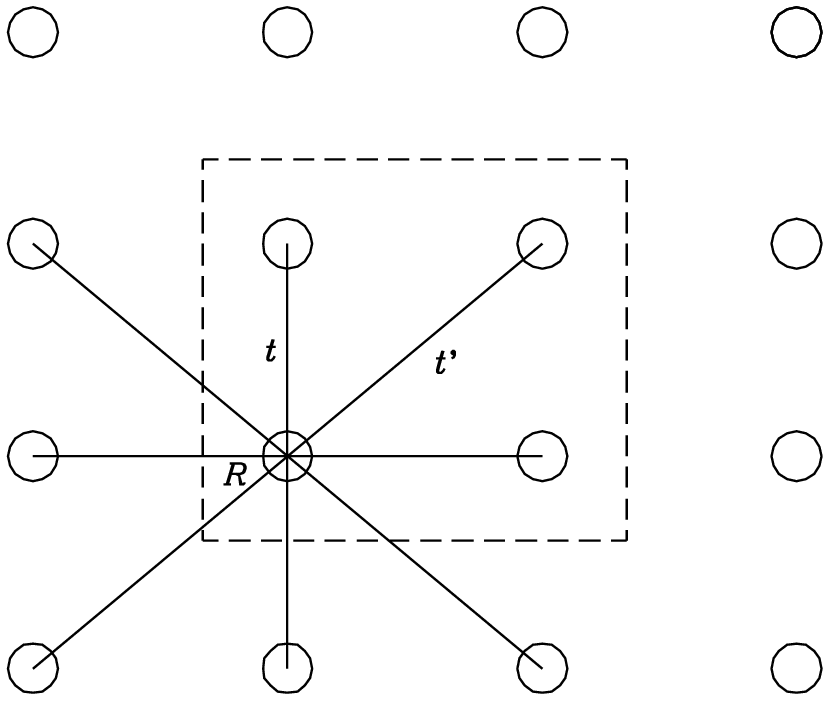}}{\fillinggrid\at(3\pscm;1\pscm)
{}}$$
\bigskip
\bigskip
\bigskip
\centerline{Fig.2:} 
$$\psannotate{\psboxto(12cm;10cm){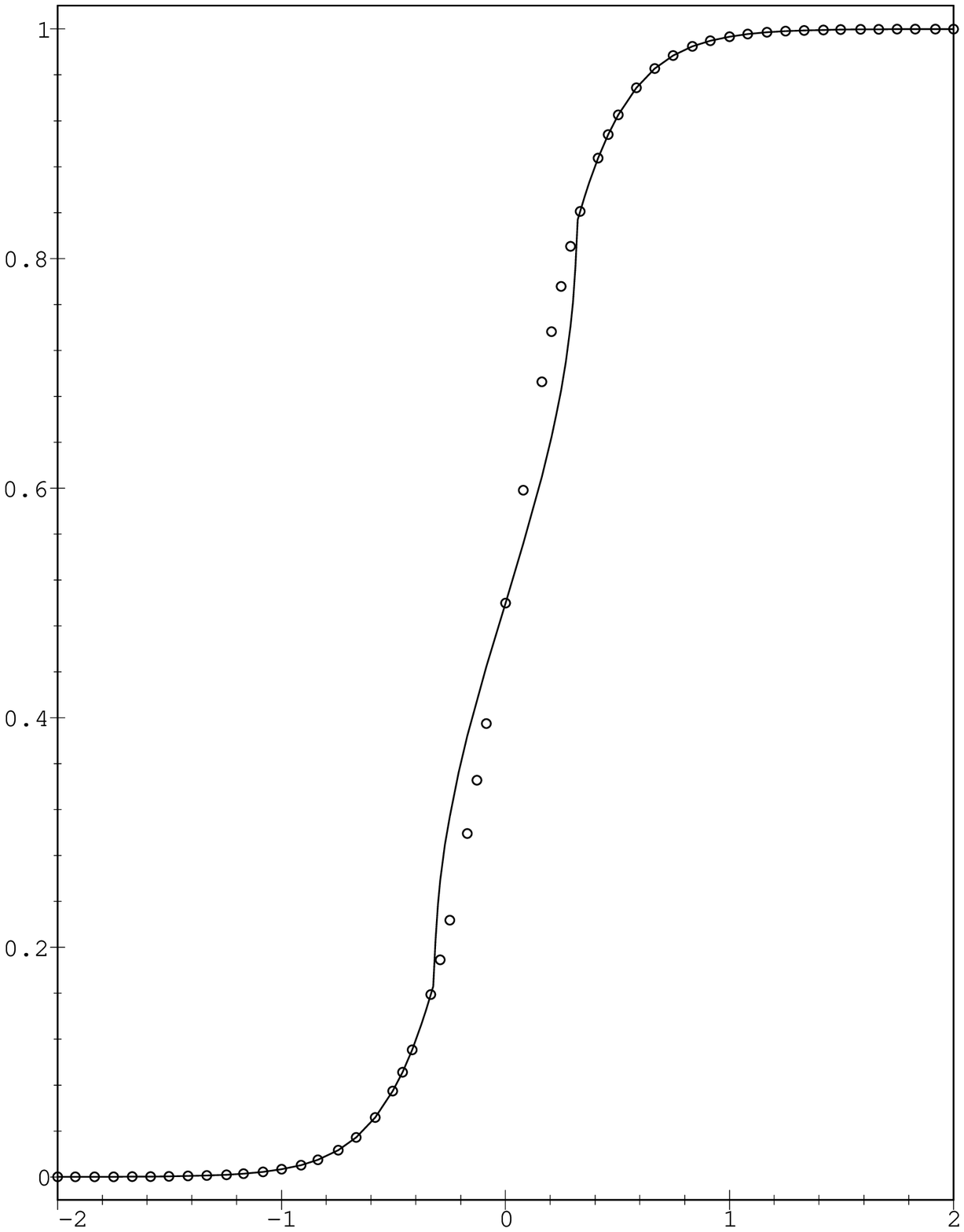}}{\fillinggrid\at(0\pscm;15\pscm)
{$\sigma_{xy}$}\at(10\pscm;2\pscm){m}}$$
\bigskip

\bye